\documentclass[%
 reprint,
 amsmath,amssymb,
 aps,
 pra,
]{revtex4-1}

\usepackage{graphicx}% Include figure files
\usepackage{dcolumn}% Align table columns on decimal point
\usepackage{bm}% bold math
\usepackage[colorlinks=true,hyperindex,breaklinks]{hyperref}% add hypertext capabilities
\begin{document}

\title{Non-Markovian dynamics of a two-level system in a bosonic bath and a Gaussian fluctuating environment with finite correlation time}

\author{V.\:A.\:Mikhailov}
 %\email{va\_mikhailov@mail.ru}
\author{N.\:V.\:Troshkin}
 %\email{nick.troshkin@gmail.com}
\affiliation{%
 Department of Physics, Samara National Research University, Moskovskoye shosse 34, 443086 Samara, Russia
}%

\date{\today}% It is always \today, today,

\begin{abstract}
The character of evolution of an open quantum system is often encoded in the correlation function of the environment or, equivalently, in the spectral density function of the interaction. When the environment is heterogeneous, e.g. consists of several independent subenvironments with different spectral functions, one of the subenvironments can be considered auxiliary and used to control decoherence of the open quantum system in the remaining part of the environment. The control can be realized, for example, by adjusting the character of interaction with the subenvironment via suitable parameters of its spectral density. We investigate non-Markovian evolution of a two-level system (qubit) under influence of three independent decoherence channels, two of them have classical nature and originate from interaction with a stochastic field, and the third is a quantum channel formed by interaction with a bosonic bath. By modifying spectral densities of the channels, we study their impact on steady states of the two-level system, evolution of its density matrix and the equilibrium emission spectrums. Additionally, we investigate the impact of the rotation-wave approximation applied to the bath channel on accuracy of the results.   
   
\end{abstract}

\maketitle

%\tableofcontents

\section{\label{sec:intro}Introduction}
In most realistic scenarios, a quantum system can not be fully isolated from its surroundings. The practically unavoidable interaction leads to emergence of decoherence processes, which are irreversible for sufficiently big environments. If the impact of these processes can not be neglected, we have to consider the quantum system as an open quantum system (OQS). When interaction with environment satisfies certain conditions, e.g. weak coupling, uncorrelated initial states, short environmental correlation times, evolution of an open quantum system (OQS) can be considered Markovian and described by a Lindblad master equation with a constant Lindblad operator and positive decay rates \cite{doi:10.1063/1.522979, doi:10.1007/BF01608499}. In many optical problems the Markovian approximation works adequately, but the restriction it imposes on the system timescales necessary for neglecting changes in the environment are too strong for modern experimental capabilities. There is a growing number of quantum systems where the memory effects associated with the environment constitute an essential part of the reduced system dynamics and can not be ignored \cite{RevModPhys.89.015001}. Among them there are such well-known systems as quantum dots \cite{PhysRevLett.106.233601,10.1126/science.aal2469}, micromechanical resonators \cite{10.1038/ncomms8606} and superconducting qubits \cite{10.1038/s41467-018-03312-x}. Many of the emerging quantum technologies, e.g. the single-photon sources for quantum communications \cite{10.1038/nphoton.2016.186}, are founded on effects induced by interaction with non-Markovian environments. Non-Markovian effects are essential for problems involving the strong interplay of vibrational and electronic states, such as electron transport in natural photosynthetic systems \cite{10.1038/nphys2515,doi:10.1146/annurev-physchem-040215-112252}, light emission in complex organic molecules and solar cells \cite{barford2013electronic}, and semiconductor quantum dots \cite{PhysRevB.84.081305,PhysRevB.83.165101,PhysRevLett.104.157401,PhysRevX.1.021009}. Many non-equilibrium quantum processes are essentially non-Markovian, like energy transport in molecular systems \cite{doi:10.1146/annurev-physchem-040215-112103} or non-adiabatic processes in physical chemistry \cite{doi:10.1146/annurev-physchem-040215-112245}. Non-Markovian evolution is also considered as a resource for quantum information tasks \cite{PhysRevA.94.052115,10.1038/srep05720,10.1209/0295-5075/107/54006}.

Usually the environment of an OQS has a large state space, and solving the full system is not feasible neither analytically nor numerically \cite{breuer2002theory,RevModPhys.89.015001}. In rare cases, exact solutions are known \cite{mahan2013many}, in others - effective weak coupling theories or perturbative expansions based on the projection operator techniques are possible \cite{PhysRevB.84.081305,doi:10.1063/1.3247899}. There are various diagrammatic and path-integral methods for which efficient Monte-Carlo schemes exist, for example, the Inchworm algorithm for the real-time diagrammatic Monte Carlo \cite{PhysRevLett.115.266802,doi:10.1063/1.4974328}. Other non-perturbative approaches include: enlarging the state space of the system \cite{PhysRevA.55.2290,PhysRevA.90.032114,schrder2017multidimensional}, capturing evolution history by augmented density tensors (ADTs) \cite{doi:10.1063/1.469508,doi:10.1063/1.469509,10.1038/s41467-018-05617-3}, mapping on effective 1-D fermionic or bosonic chains \cite{PhysRevLett.105.050404}, space reduction by thermodynamic low-dimensional approximations \cite{Semin2020}, and etc. A common property of non-perturbative methods is the strong dependence of their computational complexity on the environment correlation time. Additional restrictions on the environment type could lead to better complexity in some cases \cite{10.1143/JPSJ.58.101,schrder2017multidimensional}. One of the most widely used non-perturbative approaches is the method of hierarchical equations of motions (HEOM) \cite{10.1143/JPSJ.58.101,doi:10.1143/JPSJ.75.082001,doi:10.1063/1.4890441}. HEOMs encode the memory kernel of system-environment interaction in infinite systems of recurrent differential equations for auxiliary reduced density matrices. This approach is capable of treating a great variety of environmental spectral densities \cite{doi:10.1063/1.4766931,doi:10.1021/jz3012029,PhysRevA.85.062323,doi:10.1063/1.4870035} and works well in the high-temperature region, but at low temperatures its efficiency rapidly deteriorates due to the exponential growth of number of the Matsubara modes.     

When an open system is coupled to several environments, either classical or quantum, its measurable properties might be affected by induced correlations between the environments \cite{10.1007/s11128-009-0138-5,PhysRevA.91.022109}. If one of the environments is taken to be a stochastic environment, it becomes possible to control the system purely by noise \cite{10.1038/srep02746}. The stochastic environment alters the effective noise statistics of the joint environment in the way resembling how dynamical decoupling schemes impose effective filter functions on environmental spectral densities \cite{10.1088/0953-4075/44/15/154002,PhysRevA.90.042307} by means of ordered artificial pulses. Simultaneous action of noises of identical decoherence mechanisms has been studied in \cite{PhysRevLett.86.950,PhysRevE.69.051110}, noises of different decoherence mechanisms (dephasing and relaxation ones) in \cite{PhysRevA.91.013838}. Interplay of a bath relaxation channel and a classical noise dephasing channel for two- and three-level quantum systems is discussed in \cite{PhysRevA.97.012104,10.1016/S0375-9601(96)00805-5, Semin_2020}.  

Non-Markovian evolution differs from the Markovian one in many aspects and has its own distinctive features \cite{PhysRevA.90.042108, PhysRevA.89.012107, doi:10.1063/1.4939733, PhysRevA.99.052102}. Often it is necessary to estimate how much non-Markovianity a quantum process possesses. Many measures of non-Markovianity with specific areas of application have been developed for the purpose \cite{RevModPhys.89.015001, Rivas_2014}. Recently, several attempts have been made to measure non-Markovianity via the non-Markovian corrections for two-time correlation functions obtained via the quantum regression theorem (QRT) \cite{PhysRevA.99.012303, PhysRevA.92.062306}. Non-Markovianity can alter the QRT predictions substantially, e.g. the QRT predicts the phonon sidebands on the wrong side of the main peak in the resonance fluorescence emission spectrum for a driven semiconductor quantum dot coupled to an acoustic phonon bath \cite{PhysRevA.93.022119}. Also non-Markovianity is proved to have significant impact on equilibrium states of the reduced system \cite{PhysRevA.90.032114}.  

In the paper we investigate non-Markovian evolution of a two-level system (TLS) interacting with a composite environment that consists of an external stochastic field of arbitrary nature and a bosonic bath. The bath and the stochastic field act on the TLS independently and form one quantum decoherence channel of relaxation type and two classical decoherence channels, a pure dephasing channel and a relaxation channel. We capture non-Markovian evolution of the system exactly by utilizing the hierarchical equations of motion \cite{doi:10.1143/JPSJ.75.082001}. HEOMs are numerically exact and do not rely on any assumption, such as the strength of TLS-environment coupling. We study the evolution numerically for a stochastic field described by a set of Ornstein-Uhlenbeck random processes and a bath characterized by the high-temperature Drude spectral density.  Adjusting frequency cutoffs and coupling strengths of sub-environments, we analyze steady states of the TLS, evolution of the reduced density matrix, and equilibrium emission spectrums. 

The system Hamiltonian is often simplified with the rotating-wave approximation (RWA). The RWA neglects processes that do not conserve energy, i.e. the ones involving simultaneous creation or annihilation of energy quanta in both the TLS subspace and the environment. It is known that the RWA is able to significantly alter the entire TLS dynamics for strong TLS-bath couplings \cite{10.1088/1751-8113/43/40/405304,RevModPhys.89.015001}. For example, wrongly used RWA may lead to incorrect values for the environmentally induced shifts of the system frequencies \cite{PhysRevA.94.012110} and affects non-Markovianity properties of the evolution, as it does for interaction with bosonic environments at low temperatures \cite{PhysRevA.88.052111}. Here rapidly oscillating terms neglected by RWA mostly determine the non-Markovianity. The HOEM we developed is capable of handling both RWA and non-RWA TLS-bath couplings types equally accurate. We utilize this property to estimate the impact of the RWA on the reduced system dynamics. 

The paper is organized as follows. In Sec.~\ref{sec:model-hamiltonian} we introduce the model, in Sec.~\ref{sec:heom} we derive the hierarchical equations of motion, in Sec.~\ref{sec:markovian-me} we present the Markovian approximation. Next, we study the model numerically. In Sec.~\ref{sec:steady-states} we study steady states of a TLS, in Sec.~\ref{sec:dm-evol} we investigate evolution of the reduced density matrix, and Sec. \nolinebreak\ref{sec:spectrum} is devoted to emission spectrums. Finally, we draw conclusions in Sec. \nolinebreak\ref{sec:conclusions}.

\section{\label{sec:model-hamiltonian}Model}

The full Hamiltonian for the system can be written as
\begin{equation}
\label{HAFB}
\hat{H}=\hat{H}_A+\hat{H}_B+\hat{H}_{\text{IB}}+\hat{H}_{\text{IF}},
\end{equation}
where $\hat{H}_A=\hbar\omega_0\hat{\sigma}_+\hat{\sigma}_-$ is the Hamiltonian for the free TLS, the operators $\hat{\sigma}_{+}$ and $\hat{\sigma}_{-}$ are the rising and the lowering operators of the TLS, $\hat{H}_B=\sum_{k=1}^{\infty}\hbar\omega_k\hat{b}_k^+\hat{b}_k$ is the Hamiltonian for the free bosonic reservoir described by an infinite set of harmonic oscillators; $\hat{H}_{IF}$ is the Hamiltonian for interaction with the stochastic field 
\begin{equation}
\label{HIF}
\hat{H}_{\text{IF}}=\hbar\Omega(t)\hat{\sigma}_{+}\hat{\sigma}_{-}+\hbar[\xi(t)\hat{\sigma}_{+}+\bar{\xi}(t)\hat{\sigma}_{-}],
\end{equation}
where $\Omega(t)$, $\xi(t)$, and $\bar{\xi}(t)$ are random functions, $\Omega(t)$ is real and $\xi(t)$ is complex. The first term defines a pure dephasing channel and the second term forms a relaxation channel. We assume that all the random functions, $\Omega(t)$, $\xi(t)$, and $\bar{\xi}(t)$, are Markov processes of Ornstein-Uhlenbeck (OU) type \cite{PhysRev.36.823, fpeRisken1996, doi:10.1143/JPSJ.75.082001} and that the real and the imaginary components of $\xi(t)$ are two real OU processes $\xi_1(t)$ and $\xi_2(t)$:
\begin{align}
&\xi(t)=\xi_1(t)+i\xi_2(t),
\\
&\bar{\xi}(t)=\xi_1(t)-i\xi_2(t).
\end{align} 

The Hamiltonian $\hat{H}_{\text{IB}}$ describes interaction between the TLS and the bath. For the full electric-dipole interaction (non-RWA) it can be written
\begin{equation}
\label{HIB-non-rwa}
\hat{H}_{\text{IB}}^{\text{(full)}}=\sum _{k=1}^{\infty }g_k(\hat{\sigma}_++\hat{\sigma}_-)(\hat{b}_k^++\hat{b}_k).
\end{equation}
If the RWA is used, it takes the form
\begin{equation}
\label{HIB-RWA}
\hat{H}_{\text{IB}}^{\text{(RWA)}}=\sum _{k=1}^{\infty}(g_k\hat{\sigma}_-\hat{b}_k^{+}+\bar{g}_k\hat{\sigma}_+\hat{b}_k),
\end{equation}
where $g_k$ are the TLS-bath coupling constants. By introduction of an auxiliary TLS operator $\hat{a}$, the two forms of the interaction Hamiltonian can be combined in one expression
\begin{equation}
\label{HIB-both-in-one}
\hat{H}_{\text{IB}}=\sum _{k=1}^{\infty}(g_k\hat{a}\hat{b}_k^{+}+\bar{g}_k\hat{a}^+\hat{b}_k).
\end{equation} 
The non-RWA form of the Hamiltonian (\ref{HIB-non-rwa}) is obtained from (\ref{HIB-both-in-one}) by taking $\hat{a}=\hat{\sigma}_++\hat{\sigma}_-$, and the RWA form (\ref{HIB-RWA}) is obtained when $\hat{a}=\hat{\sigma}_-$. 

Usually, interaction with a bosonic bath is defined via the bath spectral density or the bath correlation function, which fully encodes the action of the environment on the OQS. A general form of the environmental spectral density can be written as
\begin{equation}
\label{bath-sd-general}
J(\omega)=\sum _{s=1}^{\infty}\bar{g}_s g_s \delta{(\omega - \omega_s)},
\end{equation}
Let us assume the bath is characterized by a Drude spectral density, which is the Ohmic spectral density with the algebraic cutoff \cite{doi:10.1063/1.4890441, RevModPhys.89.015001}. In the high-temperature limit it can be written as \cite{10.1143/JPSJ.58.101,doi:10.1143/JPSJ.75.082001}
\begin{equation}
\label{drude-sd}
J_D(\omega )=c_{\text{JD}}\frac{\beta\hbar\omega}{\gamma_B^2+\omega^2},
\end{equation}
where \(c_{\text{JD}}=\hbar^2\gamma_B\Delta_B^2/\pi\), the parameter \(\Delta_B\) defines the coupling strength and represents the magnitude of damping. The parameter \(\gamma_B\) stands for the width of the spectral distribution of collective bath modes and is often called a cutoff frequency. For the noise induced by the bath, $\Delta_B$ relates to the standard deviation and \(\gamma_B\) is the reciprocal of correlation time. 

\section{\label{sec:heom}Hierarchical equations of motion}

Let us introduce a basis of coherent states $|\bm{\phi}\rangle$ in the bath subspace. The basis is a cross-product of coherent states bases in each of the bath modes subspaces $|\bm{\phi}\rangle = \prod_s|\phi_s\rangle$, where s is the index of a bath mode. In the TLS subspace we introduce the basis of generalized coherent states $|z\rangle$ (also known as spin coherent states) \cite{10.1007/s10909-010-0193-4}. If we denote a vector of the full space as $|z,\bm{\phi}\rangle$, we can write the total density matrix in the 
following way
\begin{align}
\hat{\rho}&_{\text{tot}}(t)
\notag\\
&=\int_{\chi_{\Omega}}d\mu(\Omega)\int_{\chi_{\Omega}}d\mu(\Omega')
\notag\\
&\quad\times \int_{\chi_{\xi}}d\mu(\bar{\xi},\xi)
\int_{\chi_{\xi}}d\mu(\bar{\xi}',\xi')
\notag\\
&\quad\times \int_{\chi_S}d\mu(\bar{z},z)\int_{\chi_S}d\mu(\bar{z}',z')
\notag\\
&\quad\times\int_{\bm{\chi_B}}d\mu(\bm{\bar{\phi}^{(B)}},\bm{\phi^{(B)}})\int_{\bm{\chi_B}}d\mu(\bm{\bar{\phi}^{(B)}}{'},\bm{\phi^{(B)}}{'})
\notag\\
&\quad\times\rho_{\text{tot}}(\Omega,\xi,z,\bm{\phi^{(B)}},\Omega',\xi',z',\bm{\phi^{(B)}}{'},t)
\notag\\
&\quad\times|\Omega,\xi,z,\bm{\phi^{(B)}}\rangle\langle\Omega',\xi',z',\bm{\phi^{(B)}}{'}|,
\end{align}
where $\chi_\Omega$ and $\chi_\xi$ denote the sets of possible values of random processes $\Omega(t)$ and $\xi(t)$ at time $t$, respectively, $\chi_S$ denotes the TLS subspace with the infinitesimal measure
\begin{equation}
\label{tls-measure}
d\mu(\bar{z},z)=\frac{2}{(1+z\bar{z})^2}d[\bar{z},z],
\end{equation}
the integration over $\bm{\chi_B}$ is an abbreviation for integration by each of the bath modes subspaces
\begin{equation}
\int_{\bm{\chi_B}}d\mu(\bm{\bar{\phi}^{(B)}},\bm{\phi^{(B)}})=\prod_{s=1}^{\infty}\int_{\chi_B^{s}}d\mu(\bar{\phi}^{(B)s},\phi^{(B)s})
\end{equation}
with the infinitesimal measure
\begin{align}
\label{bath-measure}
d\mu(\bar{\phi}^{(B)s},\phi^{(B)s})=&\exp{(-\bar{\phi}^{(B)s}\phi^{(B)s})}
\notag\\
&\times d[\bar{\phi}^{(B)s},\phi ^{(B)s}],
\end{align}
the stochastic field subspace measures are $d\mu(\Omega)=P(\Omega)d\Omega$ and $d\mu(\bar{\xi},\xi)=P(\xi_1)P(\xi_2)d\xi_1 d\xi_2$. In (\ref{tls-measure}) and (\ref{bath-measure}) we use $d[\bar{z},z]=(1/\pi)d\,\text{Re}\,z\,d\,\text{Im}\,z$. 

Knowing the density matrix value at $t=t_0$, we can find its expression at arbitrary time moment $t>t_0$ using the evolution operators $\hat{U}(t,t_0)$ for the total system
\begin{align}
\label{total-dm-with-U}
\hat{\rho}&_{\text{tot}}(\Omega,\xi,\bar{\xi},t)
\notag\\
&=\int_{\chi_S}d\mu(\bar{z},z)\int_{\bm{\chi_B}}d\mu(\bm{\bar{\phi}^{(B)}},\bm{\phi^{(B)}})
\notag\\
&\quad\times\int_{\chi_S}d\mu(\bar{z}',z')\int_{\bm{\chi_B}}d\mu(\bm{\bar{\phi}^{(B)}}{'},\bm{\phi^{(B)}}{'})
\notag\\
&\quad\times\int_{\chi_S}d\mu(\bar{z}_0,z_0)\int_{\bm{\chi_B}}d\mu(\bm{\bar{\phi}_0^{(B)}},\bm{\phi_0^{(B)}})
\notag\\
&\quad\times\int_{\chi_S}d\mu(\bar{z}_0',z_0')\int_{\bm{\chi_B}}d\mu(\bm{\bar{\phi}_0^{(B)}}{'},\bm{\phi_0^{(B)}}{'})
\notag\\
&\quad\times U(\Omega,\xi,\bar{\xi},z,\bm{\phi^{(B)}},t;z_0,\bm{\phi_0^{(B)}},t_0)
\notag\\
&\quad\times \rho_{\text{tot}}(\Omega,\xi,\bar{\xi},z_0,\bm{\phi_0},z_0',\bm{\phi_0}',t_0)
\notag\\
&\quad\times U^+(z_0',\bm{\phi_0^{(B)}}{'},t_0;\Omega,\xi,\bar{\xi},z',\bm{\phi^{(B)}}{'},t)
\notag\\
&\quad\times|z,\bm{\phi^{(B)}}><z',\bm{\phi^{(B)}}{'}|
\end{align}

Let us suppose that the TLS was isolated from the environment before the initial moment of time and both the stochastic field and the bath were at equilibrium. Then the total density matrix at $t=t_0$, when the interaction begins, can be represented in the factorized form
\begin{equation}
\label{factorized-initial-conditions}
\hat{\rho}_{\text{tot}}(\Omega,\xi,\bar{\xi},t_0)=P_{\text{eq}}(\Omega,\xi_1,\xi_2)\hat{\rho}^{(A)}(t_0)\otimes\hat{\rho}_{\text{eq}}^{(B)}(t_0).
\end{equation}
Here the bath is taken in thermal equilibrium 
\begin{equation}
\label{rho-eq-bath}
\hat{\rho}_{\text{eq}}^{(B)}(t_0)=\exp\left(-\beta \sum_{k=1}^{\infty}\hbar\omega_{k}\hat{b}_k^{+}\hat{b}_k\right)
\end{equation}
at the inverse temperature $\beta =1/(k_B T)$, $k_B$ is the Boltzmann constant. The stochastic field state is described by the factorizable Gaussian distribution function $P_{\text{eq}}(\Omega,\xi_1,\xi_2)$ corresponding to the joint equilibrium state of the OU processes
\begin{equation}
P_{\text{eq}}(\Omega,\xi_1,\xi_2)=\prod_{\nu\in\{
\Omega,\xi_1,\xi_2\}}\frac{1}{\sqrt{2\pi\Delta_{\nu}^2}}\exp\left(-\frac{\nu^2}{2\Delta_{\nu}^2}\right),
\end{equation}  
where $\Delta_\nu$ denotes standard deviation of the random process $\nu(t)$. 

Let us divide the time interval $\left[t_0,t\right]$ on $N$ segments. At each of the time segments, evolution of the total density matrix is governed by a corresponding infinitesimal evolution operator. The total evolution operator $\hat{U}(t_0,t)$ is a product of the infinitesimal operators for each of the time segments. By inserting $N-1$ identity operators at respective $N-1$ moments of time and taking the limit $N\to\infty$, we obtain matrix elements of the total evolution operator in the next form
\begin{align}
\label{evolution-operator-matrix-elements}
U&(\Omega,\xi,z,\bm{\phi^{(B)}},t;\Omega_0,\xi_0,z_0,\bm{\phi_0^{(B)}},t_0)
\notag\\
&=\int_{\mathcal{C}\{\Omega,t;\Omega_0,t_0\}}\mathcal{D}[\Omega(\tau)]
\int_{\mathcal{C}\{\xi,t;\xi_0,t_0\}}\mathcal{D}[\xi(\tau)] 
\notag\\
&\quad\times \int_{\mathcal{C}\{z,t;z_0,t_0\}}\mathcal{D}[\bar{z}(\tau),z(\tau)]
\notag\\
&\quad\times\int_{\mathcal{C}\{\bm{\phi^{(B)}},t;\bm{\phi_0^{(B)}},t_0\}}\mathcal{D}[\bm{\bar{\phi}^{(B)}}(\tau),\bm{\phi^{(B)}}(\tau)]
\notag\\
&\quad\times P[\Omega(\tau)]P[\xi(\tau)]\, 
\notag\\
&\quad\times \exp\left(\frac{i}{\hbar}S_A[z;t,t_0]+\frac{i}{\hbar}S_B[\bm{\phi^{(B)}};t,t_0]\right.
\notag\\
&\qquad-\frac{i}{\hbar}\int_{t_0}^t d\tau H_{\text{IF}}\bm(\bar{z}(\tau);z(\tau),\xi(\tau),\bar{\xi}(\tau),\Omega(\tau)\bm)
\notag\\
&\qquad\left.-\frac{i}{\hbar}\int_{t_0}^t d\tau H_{\text{IB}}\bm(\bar{z}(\tau),\bm{\bar{\phi}^{(B)}}(\tau);z(\tau),\bm{\phi^{(B)}}(\tau)\bm)\right),
\end{align}
where $\int\mathcal{D}[\bar{z}(\tau),z(\tau)]$ denotes functional integration over the set of trajectories starting at $z(t_0)=z_0$ and ending at $z(t)=z_N=z$,
\begin{equation}
\int_{\mathcal{C}\{z,t;z_0,t_0\}}\mathcal{D}[\bar{z}(\tau ),z(\tau )]=\lim_{N\to\infty}\prod
_{j=1}^{N-1}\int_{\chi_S}d[\bar{z}_j,z_j],
\end{equation}
and $\int\mathcal{D}[\bm{\bar{\phi}^{(B)}}(\tau),\bm{\phi^{(B)}}(\tau)]$ denotes path integrals over the set of trajectories of all the bath modes,
\begin{align}
\int&_{\mathcal{C}\{\bm{\phi^{(B)}},t;\bm{\phi_0^{(B)}},t_0\}}\mathcal{D}[\bm{\bar{\phi}^{(B)}}(\tau ),\bm{\phi^{(B)}}(\tau )]
\notag\\
&=\lim_{N\to\infty}\prod_{s=1}^{\infty }\prod_{j=0}^{N-1}\int_{\chi_B}d[\bar{\phi}_j^{(B)s},\phi_j^{(B)s}],
\end{align}
$\int\mathcal{D}[\Omega(\tau)]$ and $\int\mathcal{D}[\xi(\tau)]=\int\mathcal{D}[\xi_1(\tau)]\int\mathcal{D}[\xi_2(\tau)]$ denote Wiener-type path integrals over realizations of the stochastic field, $P[\Omega(\tau)]$ and $P[\xi(\tau)]=P[\xi_1(\tau)]P[\xi_2(\tau)]$ are probability functionals defining probabilities of stochastic trajectories of the respective OU processes.

The functional $S_A[z;t,t_0]$ denotes the action for the free TLS and has the next form  
\begin{align}
S&_A[z;t,t_0]
\notag\\
&=-i \hbar\lim_{N\to\infty}\sum_{j=0}^{N-1}\{-\ln[(1+z_j \bar{z}_j)^2/2]+\ln(\langle z_{j+1}|z_j\rangle)
\notag\\
&\quad-\epsilon(i/\hbar)H_A(z_{j+1},z_j)\},
\end{align}
the functional $S_B[\bm{\phi^{(B)}};t,t_0]$ is the action for the bath Hamiltonian
\begin{align}
S&_B[\bm{\phi^{(B)}};t,t_0]
\notag\\
&=-i\hbar\lim_{N\to\infty}\sum_{j=0}^{N-1}
[\bm{\bar{\phi}_{j+1}^{(B)}}\bm{\phi_j^{(B)}}-\bm{\bar{\phi}_j^{(B)}}\bm{\phi_j^{(B)}}
\notag\\
&\quad-\frac{i}{\hbar}\epsilon\,H_B(\bm{\phi_{j+1}^{(B)}},\bm{\phi_j^{(B)}})],
\end{align}
where $\epsilon=(t-t_0)/N$ is the length of a time segment, $H_A$, $H_B$, $H_{\text{IF}}$, and $H_{\text{IB}}$ are symbols of $\hat{H}_A$, $\hat{H}_B$, $\hat{H}_{\text{IF}}$, and $\hat{H}_{\text{IB}}$ operators, respectively. All variables having a $j$ index correspond to the $j$-th time slice or the $j$-th identity operator. The semicolons in arguments of the operator symbols separate variables taken at different slices. Those to the left correspond to the one-step-forward slice, with index $j+1$, and to the right from a semicolon - to the $j$-the slice. 

Tracing out the bath and the stochastic field degrees of freedom, we obtain the reduced density matrix for the TLS subsystem with the following matrix elements
\begin{align}
\label{averaged-atom-density-matrix-elements}
\rho&^{(A)}(z,z',t)
\notag\\
&=\int_{\mathcal{C}\{z,t;\chi_S,t_0\}}\mathcal{D}[\bar{z}(\tau),z(\tau)]
\int_{\mathcal{C}\{z',t;\chi_S,t_0\}}\mathcal{D}[\bar{z}'(\tau),z'(\tau)]
\notag\\
&\quad\times \rho^{(A)}(z_0,z_0',t_0)
\notag\\
&\quad\times F_F[z(\tau),z'(\tau);t,t_0]
F_B[z(\tau),z'(\tau);t,t_0]
\notag\\
&\quad\times \exp\left(\frac{i}{\hbar}S_A[z;t,t_0]-\frac{i}{\hbar}\bar{S}_A[z';t,t_0]\right),
\end{align}
where we have utilized the form of the evolution operator (\ref{evolution-operator-matrix-elements}) and the factorized from of the initial conditions (\ref{factorized-initial-conditions}), $F_F[z(\tau),z'(\tau);t,t_0]$ is the stochastic field influence functional that is a product of influence functionals for each of the random processes
\begin{equation}
\label{total-field-influence-functional}
F_F[z(\tau),z'(\tau);t,t_0]=\Pi_{\nu}F_{\nu }[z(\tau),z'(\tau);t,t_0],
\end{equation}
where $\nu\in\{\Omega,\xi_1,\xi_2\}$ and $F_{\nu}[z(\tau),z'(\tau);t,t_0]$ is represented by a functional integral of the next form 
\begin{align}
\label{random-process-influence-functional}
F&_{\nu}[z(\tau),z'(\tau);t,t_0]
\notag\\
&=\int_{\mathcal{C}\{\chi_{\nu},t;\chi_{\nu },t_0\}}\mathcal{D}[\bar{\nu}(\tau),\nu(\tau)]
P[\nu(\tau)]P_{\text{eq}}(\nu_0)
\notag\\
&\quad\times \exp\left(-\frac{i}{\hbar}\int_{t_0}^t d\tau H_{I,\nu}\bm(\bar{z}(\tau);z(\tau),\nu(\tau)\bm)\right.
\notag\\
&\qquad+\left.\frac{i}{\hbar}\int_{t_0}^t d\tau \bar{H}_{I,\nu}\bm(\bar{z}'(\tau);z'(\tau),\nu(\tau)\bm)\right)
\end{align} 
and $F_B[z(\tau),z'(\tau);t,t_0]$ denotes the bath influence functional
\begin{align}
\label{FB}
F&_B[z(\tau),z'(\tau);t,t_0]
\notag\\
&=\int_{\mathcal{C}\{\bm{\chi_B},t;\bm{\chi_B},t_0\}}\mathcal{D}[\bm{\bar{\phi}^{(B)}}(\tau),\bm{\phi^{(B)}}(\tau)]
\notag\\
&\quad\times \int_{\mathcal{C}\{\bm{\chi_B},t;\bm{\chi_B},t_0\}}\mathcal{D}[\bm{\bar{\phi}^{(B)}}{'}(\tau),\bm{\phi^{(B)}}{'}(\tau )]
\notag\\
&\quad\times \exp\left(\frac{i}{\hbar}S_B[\bm{\phi^{(B)}};t,t_0]\right.
\notag\\
&\qquad\left.-\frac{i}{\hbar}\int_{t_0}^t d\tau H_{\text{IB}}\bm(\bar{z}(\tau),\bm{\bar{\phi}^{(B)}}(\tau);z(\tau),\bm{\phi^{(B)}}(\tau)\bm)\right)
\notag\\
&\quad\times e^{\bm{\bar{\phi}^{(B)}}{'}(t)\bm{\phi^{(B)}}(t)}
\rho_{\text{eq}}^{(B)}(\bm{\phi_0^{(B)}},\bm{\phi_0^{(B)}}{'},t_0)
\notag\\
&\quad\times \exp\left(-\frac{i}{\hbar}\bar{S}_B[\bm{\phi^{(B)}}{'};t,t_0]\right.
\notag\\
&\qquad\left.+\frac{i}{\hbar}\int_{t_0}^t d\tau \bar{H}_{\text{IB}}\bm(\bar{z}'(\tau),\bm{\bar{\phi}^{(B)}}{'}(\tau);z'(\tau),\bm{\phi^{(B)}}{'}(\tau)\bm)\right).
\end{align}

The product form of the field influence functional (\ref{total-field-influence-functional}) is a consequence of absence of initial correlations between the TLS and the stochastic field and mutual independence of the random processes $\xi_1(t)$, $\xi_2(t)$, and $\Omega(t)$. In (\ref{random-process-influence-functional}) we have introduced symbols of interaction Hamiltonians with each of the random processes
\begin{equation}
H_{I,\nu}(z_{j+1};z_j,\nu_j)=\nu_j V_{F,\nu}(z_{j+1};z_j),
\end{equation}
where $V_{\nu}$ are symbols of the operators, defined on the TLS subspace
\begin{align}
\label{VF1}
\hat{V}_{F,\Omega}&=\hbar \hat{J_0},
\\
\hat{V}_{F,\xi_1}&=\hbar (\hat{\sigma}_++\hat{\sigma}_-),
\\
\label{VF2}
\hat{V}_{F,\xi_2}&=i \hbar (\hat{\sigma}_+-\hat{\sigma}_-).
\end{align}

The hierarchical equations of motion (HEOM) are obtained from the reduced density matrix (\ref{averaged-atom-density-matrix-elements}) by repetitive differentiation of the memory kernel, related to the influence functional \cite{10.1143/JPSJ.58.101,doi:10.1143/JPSJ.75.082001}. At first we consider a simplified problem with the stochastic field turned off. From the two interaction Hamiltonians in Eq.(\ref{HAFB}) we keep only $\hat{H}_{\text{IB}}$, so the expression (\ref{averaged-atom-density-matrix-elements}) for the averaged reduced density matrix has only one influence functional $F_B[z(\tau),z'(\tau);t,t_0]$. Increasing t by a small value $\epsilon$, we get
\begin{align}
\label{rhoAt+e-definition}
\rho&^{(A)}(z,z',t+\epsilon)
\notag\\
&=\int_{\mathcal{C}\{z,t+\epsilon;\chi_S,t_0\}}\mathcal{D}[\bar{z}(\tau),z(\tau)]
\notag\\
&\quad\times \int_{\mathcal{C}\{z',t+\epsilon;\chi_S,t_0\}}\mathcal{D}[\bar{z}'(\tau),z'(\tau)]
\notag\\
&\quad\times \rho ^{(A)}(z_0,z_0',t_0)
F_B[z(\tau),z'(\tau);t+\epsilon ,t_0]
\notag\\
&\quad\times \exp\left(\frac{i}{\hbar}S_A[z;t+\epsilon,t_0]-\frac{i}{\hbar}\bar{S}_A[z';t+\epsilon,t_0]\right).
\end{align}
Taking values at $t+\epsilon$ involves one extra segment on the time axis, lying on the right from $t$ and of length $\epsilon$. Thus, the increment
of the free-TLS action can be written as 
\begin{align}
\label{SAt+e}
S&_A[z;t+\epsilon,t_0]
\notag\\
&=S_A[z;t,t_0]+i\hbar\ln[(1+z_N \bar{z}_N)^2/2]-i\hbar\ln(\langle z|z_N\rangle)
\notag\\
&\quad-\epsilon H_A(z,z_N).
\end{align}
The influence functional $F_B[z(\tau),z'(\tau);t,t_0]$ can be found from the discreet form of the path integral (\ref{FB}) by performing the bosonic Gaussian integration. For the relaxation-type interaction Hamiltonian (\ref{HIB-both-in-one}) the following expression can be obtained 
\begin{align}
\label{FB-after-gaussian-integration}
F&_B[z(\tau),z'(\tau);t,t_0]
\notag\\
&=\exp\left[\left(-\frac{i}{\hbar}\right)^2
\sum_{k=1}^2 
\int_{t_0}^{t}dt'\,\Phi_k[z(\tau),z'(\tau);t']V_k^{\text{x}}(t')\right],
\end{align}
where we have introduced the influence phase functionals
\begin{equation}
\label{PhiVi}
\Phi_k[z(\tau),z'(\tau);t]=\int_{t_0}^t dt' \int_0^{+\infty}d\omega J(\omega)f_k(t,t',\omega),
\end{equation}
with
\begin{align}
\label{fV1}
&f_1(t,t',\omega)\nonumber
\\
&\quad=e^{i\omega (t-t')}
[n_B(\omega)C_2^+(t')-(n_B(\omega)+1)C_2^-(t')],
\\
&f_2(t,t',\omega)\nonumber
\\
\label{fV2}
&\quad=e^{-i \omega (t-t')}
[(n_B(\omega)+1)C_1^+(t')-n_B(\omega)C_1^-(t')].
\end{align}
Here, $n_B(\omega)=\rho(\omega)/[1-\rho(\omega)]$ and $\rho(\omega)=\exp{(-\beta \hbar \omega)}$, $C_1^{\pm}(t)$, $C_2^{\pm}(t)$ are continuous representation of symbols of the auxiliary operators $\hat{a}$ and $\hat{a}^+$ respectively, the plus and the minus signs originate from the forward and the backward branches of the path integral; $V_1^{\text{x}}(\tau)=C_1^+(t)-C_1^-(t)$, $V_2^{\text{x}}(\tau)=C_2^+(t)-C_2^-(t)$, and $J(\omega)$ is the bath spectral density (\ref{bath-sd-general}).

In case of the non-RWA interaction with the bath (\ref{HIB-non-rwa}), the equation (\ref{FB-after-gaussian-integration}) can be transformed to the form of the well-known influence functional of Feynman and Vernon \cite{10.1016/0003-4916(63)90068-X}. If the RWA approximation has been applied, some symmetry of the equation has been lost with the neglected terms, and the equation cannot be represented in this form, in contrast with the non-RWA case.

Incrementing the time argument of $F_B[z(\tau),z'(\tau);t,t_0]$, we get 
\begin{align}
\label{FBt+e}
F&_B[z(\tau),z'(\tau);t+\epsilon,t_0]
\notag\\
&=\left(1+\epsilon \sum_{k=1}^2\Phi_{B,k}^{(0)}(t+\epsilon)\Phi_k[z(\tau),z'(\tau);t]\right)
\notag\\
&\quad\times F_B[z(\tau),z'(\tau);t,t_0],
\end{align}
where the functions of the boundary time
\begin{equation}
\Phi_{B,k}^{(0)}(t+\epsilon)=(-i/\hbar)^2 V_k^{\text{x}}(t+\epsilon)
\end{equation}
depend on the previous time slice variables $z_N$ and $z_N'$

Substituting (\ref{SAt+e}) and (\ref{FBt+e}) into (\ref{rhoAt+e-definition}) and introducing auxiliary density matrices forming the hierarchy of the bath memory kernel, we obtain the expression for the increment of the reduced density matrix allowing the inverse transformation to the operator form  
\begin{align}
\rho&^{(A)}(z,z',t+\epsilon)-\rho^{(A)}(z,z',t)
\notag\\
&=\,\epsilon\int_{\chi_S}d\mu(\bar{z}_N,z_N)\langle z|z_N\rangle
\int_{\chi_S}d\mu(\bar{z}_N',z_N')\langle z_N'|z'\rangle
\notag\\
&\quad\times \left(-\frac{i}{\hbar}[H_A(z,z_N)-H_A(z_N',z')]\rho^{(A)}(z_N,z_N',t)\right.
\notag\\
&\qquad+\Phi_{B,k}^{(0)}(t+\epsilon)\sum_{k=1}^2\rho_k^{(A)}(z_N,z_N',t)\bigg),
\end{align}
where the auxiliary density matrices are 
\begin{align}
\label{rho1-definition}
\rho&_k^{(A)}(z_N,z_N',t)
\notag\\
&=\int_{\mathcal{C}\{z_N,t;\chi_S,t_0\}}\mathcal{D}[\bar{z}(\tau),z(\tau)]
\notag\\
&\quad\times \int_{\mathcal{C}\{z_N',t;\chi_S,t_0\}}\mathcal{D}[\bar{z}'(\tau),z'(\tau)]
\notag\\
&\quad\times \Phi_k[z(\tau),z'(\tau);t]\,F_B[z(\tau),z'(\tau);t,t_0]\rho^{(A)}(z_0,z_0',t_0)
\notag\\
&\quad\times \exp\left(\frac{i}{\hbar}S_A[z;t,t_0]-\frac{i}{\hbar}\bar{S}_A[z';t,t_0]\right).
\end{align}

It can be seen from (\ref{rho1-definition}) that, because $\Phi_1[z(\tau),z'(\tau);t]$ is not equal to $\Phi_2[z(\tau),z'(\tau);t]$, there are two branches in the recursion relation for the bath. The time-incremented form of (\ref{rho1-definition}) involves $\Phi_1[z(\tau),z'(\tau);t+\epsilon]$ and $\Phi_2[z(\tau),z'(\tau),t+\epsilon]$, which can be obtained from (\ref{PhiVi}) by incrementing the time argument
\begin{align}
\label{Phi-k-increment}
\Phi_k[z(\tau),z'(\tau);t+\epsilon]=&\Phi_k[z(\tau),z'(\tau);t]
\notag\\
&+\epsilon\Phi_{B,k}^{(1)}(t+\epsilon)
\notag\\
&+\epsilon\Psi_{B,k}^{(1)}[z(\tau),z'(\tau);t],
\end{align}
where
\begin{equation}
\label{deltaPhiPoint-def}
\Phi_{B,k}^{(1)}(t+\epsilon)=\frac{1}{2}\int_{-\infty}^{+\infty}d\omega J(\omega)f_k(t+\epsilon,t+\epsilon,\omega)
\end{equation}
and
\begin{align}
\label{delta-phi-path-def}
\Psi_{B,k}^{(1)}[z(\tau),z'(\tau);t]=&\frac{1}{2}\int_0^t dt' \int_{-\infty}^{+\infty}d\omega J(\omega)
\notag\\
&\times \frac{\partial}{\partial t}f_k(t,t',\omega).
\end{align}
In (\ref{deltaPhiPoint-def}) and (\ref{delta-phi-path-def}) the lower limits of integration by $\omega$ have been extended to the negative infinity. The limits can be extended, if $\hat{a}$ is a self-conjugated operator and the spectral density $J(\omega)$ is an odd function of $\omega$. It is completely fulfilled for the non-RWA interaction, but in the RWA case, $\hat{a}$ is not a self-conjugated operator, and the extension imposes rather strong restrictions on both the coupling strength and the maximum time of the dynamics.

In (\ref{Phi-k-increment}), $\Phi_{B,k}^{(1)}(t+\epsilon)$ depends only on the boundary time and yields an operator in the operator form of HEOM, while $\Psi_{B,k}^{(1)}[z(\tau),z'(\tau);t]$ still depends on the whole path and contains the memory of the interaction. For the environments considered in the current work, $\Psi_{B,k}^{(1)}[z(\tau),z'(\tau);t]$ satisfies the following relation
\begin{equation}
\label{Phi-path-recursion-condition}
\Psi_{B,k}^{(1)}[z(\tau),z'(\tau);t]=\alpha_k^{(B)}\Phi_k[z(\tau),z'(\tau);t],
\end{equation}
where $\alpha_k$ is a constant.

The relation (\ref{Phi-path-recursion-condition}) yields a system of ODEs for the auxiliary density matrices that can be translated to the operator form. The system of operator ODEs is of the HEOM type and can be written as 
\begin{align}
\label{atom-bath-heom}
\frac{\partial}{\partial t}\hat{\rho}_{m,n}^{(A)}(t)=&[-(i/\hbar) \hat{H}_A^{\text{x}}+m \alpha_1^{(B)}+n \alpha_2^{(B)}]\hat{\rho}_{m,n}^{(A)}(t)
\notag\\
&+\hat{\Phi}_{B,1}^{(0)}\hat{\rho}_{m+1,n}^{(A)}(t)+\hat{\Phi}_{B,2}^{(0)}\hat{\rho}_{m,n+1}^{(A)}(t)
\notag\\
&+m\,\hat{\Phi}_{B,1}^{(1)}\hat{\rho}_{m-1,n}^{(A)}(t)+n\,\hat{\Phi}_{B,2}^{(1)}\hat{\rho}_{m,n-1}^{(A)}(t),
\end{align}
where we split the index of (\ref{rho1-definition}) on two, by one for each of the branches; by \(\hat{H}_A^{\text{x}}\) we denote the commutator super-operator \(\hat{H}_A^{\text{x}}\hat{\rho }=\hat{H}_A^{\text{x}}\hat{\rho}-\hat{\rho}\hat{H}_A^{\text{x}}\).

In (\ref{atom-bath-heom}) only $\hat{\rho}_{0,0}^{(A)}(t)$ has a physical meaning, the others constitutes the bath memory kernel representation. By redefining the auxiliary memory functions (\ref{rho1-definition}), we can obtain an equivalent HEOM with adjusted coefficients, e.g. by making the substitutions $\hat{\rho}_{m+1,n}^{(A)}(t) \to a\hat{\rho}_{m+1,n}^{(A)}(t)$ and $\hat{\rho}_{m-1,n}^{(A)}(t) \to (1/a)\hat{\rho}_{m-1,n}^{(A)}(t)$.

Similarly we perform the Gaussian integration in a Wiener path integral \cite{doi:10.1143/JPSJ.75.082001} and obtain a HEOM resembling eq. (\ref{atom-bath-heom}) with three indexes for interaction with the stochastic field. For the joint environment consisting of the HT-Drude bath and the stochastic field, a HEOM with five indexes may be obtained. Let us introduce the vector notation for indexes of the auxiliary density matrices 
\begin{align*}
\bm{m}&=(m_1,m_2,\dotsc),
\\
\bm{m|_{k+1}}&=(m_1,m_2,\dotsc,m_k+1,\dotsc).
\end{align*}
By means of it the HEOM can be written in the following form
\begin{align}
\label{atom-bath-field-heom}
\frac{\partial }{\partial t}\hat{\rho}_{\bm{m}}^{(A)}(t)=&-\frac{i}{\hbar} \hat{H}_A^{\text{x}}\hat{\rho}_{\bm{m}}^{(A)}(t)
\notag\\
&+\sum_{k\in\{\text{field}\}}[m_k \alpha_k^{(F)}\hat{\rho}_{\bm{m}}^{(A)}(t)+\hat{\Phi}_{F,k}^{(0)}\hat{\rho}_{\bm{m|_{k+1}}}^{(A)}(t)
\notag\\
&+m_k\hat{\Phi}_{F,k}^{(1)}\hat{\rho}_{\bm{m|_{k-1}}}^{(A)}(t)]
\notag\\
&+\sum_{k\in\{\text{bath}\}}[m_k \alpha_{k}^{(B)}\hat{\rho}_{\bm{m}}^{(A)}(t)+ \hat{\Phi}_{B,k}^{(0)}\hat{\rho}_{\bm{m|_{k+1}}}^{(A)}(t)
\notag\\
&+m_k\hat{\Phi}_{B,k}^{(1)}\hat{\rho}_{\bm{m|_{k-1}}}^{(A)}(t)],
\end{align}
where $\hat{\rho }^{(A)}(t)=\hat{\rho}_{\bm{0}}(t)$ and we make the summations over the field and the bath indexes explicit. 

The condition (\ref{Phi-path-recursion-condition}) is satisfied for Ornstein-Uhlenbeck stochastic fields, for which the unknown constant \(\alpha_k^{(F)}\) and the unknown operators \(\hat{\Phi}_{F,k}^{(0)}\) and \(\hat{\Phi}_{F,k}^{(1)}\) are
\begin{align}
\alpha_k^{(F)}=&-\gamma_{\nu_k},
\\
\hat{\Phi}_{F,k}^{(0)}=&-\Delta_{\nu_k}(i/\hbar)\hat{V}_{F,\nu_k}^{\text{x}},
\\
\hat{\Phi}_{F,k}^{(1)}=&-\Delta_{\nu_k}(i/\hbar)\hat{V}_{F,\nu_k}^{\text{x}},
\end{align}
where $\nu_k$ corresponds to the k-th element of $\{\Omega,\xi_1,\xi_2\}$, $\hat{V}_{F,\nu_k}^{\text{x}}\hat{\rho}=\hat{V}_{F,\nu_k}\hat{\rho }-\hat{\rho }\hat{V}_{F,\nu_k}$, and $\hat{V}_{F,\nu_k}$ are defined in (\ref{VF1}-\ref{VF2}). Parameters $\gamma_{\nu_k}$ and $\Delta_{\nu_k}$ denote the reciprocal of correlation time and the standard deviation of the random process $\nu_k(t)$.

The relation (\ref{Phi-path-recursion-condition}) is also satisfied for the high-temperature Drude bath (\ref{drude-sd}). In the high-temperature limit, \(n_B(\omega)\) in Eqs. (\ref{fV1}, \ref{fV2}) contains a small parameter $\beta\hbar\omega$. Because $\omega$ is bounded by the effective cutoff frequency, which is defined by \(\gamma_B\), the validity of the approximation is restricted by $\beta\hbar\gamma_B\ll\nolinebreak1$. In the region of validity, the HEOM coefficients take the next form
\begin{align}
\label{drude-sd-heom-coeffs-first}
\alpha_{k}^{(B)}&=-\gamma_B,
\\
\hat{\Phi}_{B,k}^{(0)}&=a_B(-i/\hbar)^2\hat{c}_k^{\text{x}},
\\
\hat{\Phi}_{B,1}^{(1)}&=\frac{1}{a_B}\frac{c_{\text{JD}}}{2}\frac{\pi}{\gamma _B}(\hat{c}_2^{\text{x}}-i \beta \hbar \gamma_B\hat{c}_2),
\\
\label{drude-sd-heom-coeffs-last}
\hat{\Phi}_{B,2}^{(1)}&=\frac{1}{a_B}\frac{c_{\text{JD}}}{2}\frac{\pi}{\gamma_B}(\hat{c}_1^{\text{x}}-i \beta\hbar\gamma_B\hat{c}_1^R),
\end{align}
where \(\hat{c}_1=\hat{a}\), \(\hat{c}_2=\hat{a}^+\), and \(\hat{c}_k^R\) is the super-operator acting from the right, \(\hat{c}_k^R\hat{\rho }=\hat{\rho
}\hat{c}_k^R\). It is also convenient to use the renormalization constant \(a_B=\hbar^2\Delta
_B\) for better HEOM coefficients.

For the non-RWA interaction with a high-temperature Drude bath, it is possible to obtain a one-index HEOM \cite{10.1143/JPSJ.58.101}. This HEOM resembles the one for the stochastic field when there is only one random process is present, and the closeness increases with temperature  \cite{doi:10.1143/JPSJ.75.082001}. For the RWA interaction the minimum number of HEOM indexes is two because of the ambiguity in the transition from the path integral form to the operator form, also known as the quantization problem.   

\section{\label{sec:markovian-me}Markovian approximation}
In the Markovian approximation, evolution of the TLS reduced density matrix is described by the Markovian master equation of Lindblad type \cite{breuer2002theory}
\begin{equation}
\label{markovian-me}
\frac{d}{dt}\hat{\rho}_{\text{int}}^{(A)}=-\frac{i}{\hbar}[\hat{H}_{\text{LS}},\hat{\rho}_{\text{int}}^{(A)}]+\mathcal{D}_B(\hat{\rho}_{\text{int}}^{(A)})
+\mathcal{D}_F(\hat{\rho}_{\text{int}}^{(A)}),
\end{equation}
where \(\hat{\rho}_{\text{int}}^{(A)}\) stands for the density matrix in the interaction picture, \(\hat{H}_{\text{LS}}\) is the so called Lamb shift Hamiltonian 
\begin{equation}
\label{Lamb-shift-Hamiltonian}
\hat{H}_{\text{LS}}=\hbar S_{\text{LS}}(\omega_0)\hat{\sigma}_+\hat{\sigma}_-+\hbar S_{\text{LS}}(-\omega_0)\hat{\sigma}_-\hat{\sigma}_+,
\end{equation}
with \(S_{\text{LS}}(\omega_0)\) depending on the spectral density type. For the high-temperature limit of the Drude spectral density it has the form
\begin{equation}
S_{\text{LS}}(\omega )=c_{\text{JD}}\frac{2\pi\omega+\hbar\beta\gamma_B[2 \omega\ln(\gamma_B/|\omega|)-\pi\gamma_B]}{2\gamma_B (\gamma_B^2+\omega^2)}.
\end{equation}
The symbol \(\mathcal{D}_B(\hat{\rho }_{\text{int}}^{(A)})\) denotes the bath-related dissipator 
\begin{align}
\mathcal{D}_B(\hat{\rho}_{\text{int}}^{(A)})=&\frac{2\pi}{\hbar^2} J(\omega_0)(n_B(\omega_0)+1)
\notag\\
&\quad\times(\hat{\sigma}_-\hat{\rho}_{\text{int}}^{(A)}\hat{\sigma}_+-\frac{1}{2}\{\hat{\sigma}_+\hat{\sigma}_-,\hat{\rho}_{\text{int}}^{(A)}\})
\notag\\
&+\frac{2\pi}{\hbar^2} J(\omega_0)n_B(\omega_0)
\notag\\
&\quad\times(\hat{\sigma}_+\hat{\rho}_{\text{int}}^{(A)}\hat{\sigma}_--\frac{1}{2}\{\hat{\sigma}_-\hat{\sigma}_+,\hat{\rho }_{\text{\text{int}}}^{(A)}\}),
\end{align}
and \(\mathcal{D}_F(\hat{\rho }_{\text{int}}^{(A)})\) is the stochastic field dissipator \cite{mikhailov2016master} 
\begin{align}
\mathcal{D}_F(\hat{\rho }_{\text{int}}^{(A)})
=&2K_{\Omega}(t)(\hat{J}_0\hat{\rho }_{\text{int}}^{(A)}\hat{J}_0-\frac{1}{2}\{\hat{J}_0^2,\hat{\rho}_{\text{int}}^{(A)}\})
\notag\\
&+2K_{\xi}(t)(\hat{\sigma}_-\hat{\rho }_{\text{int}}^{(A)}\hat{\sigma}_+-\frac{1}{2}\{\hat{\sigma}_+\hat{\sigma}_-,\hat{\rho}_{\text{int}}^{(A)}\})
\notag\\
&+2K_{\xi}(t)(\hat{\sigma}_+\hat{\rho }_{\text{int}}^{(A)}\hat{\sigma}_--\frac{1}{2}\{\hat{\sigma}_-\hat{\sigma}_+,\hat{\rho
}_{\text{int}}^{(A)}\}),
\end{align}
where the curl brackets mean anticommutators of the operators, \(K_{\Omega}(t)\) and \(K_{\xi}(t)\) are integrals of correlation functions of corresponding random processes, \(K_{\nu}(t)=\int_{t_0}^t\langle \nu(t)\bar{\nu}(t_1)\rangle dt_1\).

\section{\label{sec:steady-states}Steady states}
\begin{figure*}[t]
\includegraphics[width=\linewidth]{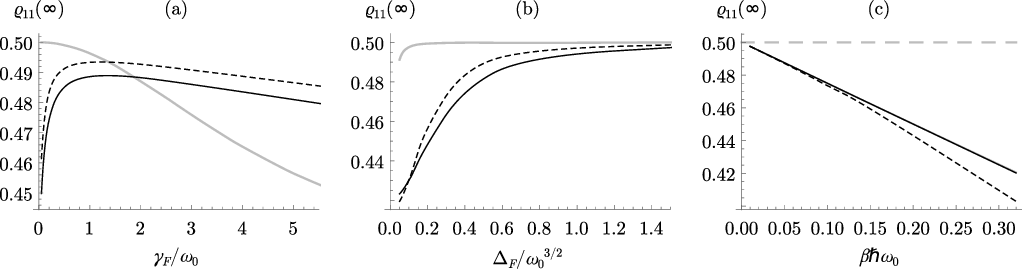}
\caption{
\label{fig:steady-states}
(a,b) The TLS excited state population in equilibrium for simultaneous interaction with the stochastic field and the HT-Drude bath as a function of (a) the field frequency cutoff $\gamma_F$ and (b) the field coupling strength $\Delta_F$. The dashed and the solid black curves denote the RWA and the non-RWA couplings, respectively, the gray curve stands for the Markovian approximation. In (a) $\gamma_\nu=\gamma_F$ and $\Delta_\nu/\omega_0^{3/2}=0.4$, in (b) $\gamma_\nu/\omega_0=0.2$ and $\Delta_\nu=\Delta_F$, the bath is the same, $\gamma_B/\omega_0=0.2$, $\Delta_B/\omega_0^{1/2}=0.4$, $\beta\hbar\omega_0=0.32$.
	(c) The TLS excited state population in equilibrium for interaction with the HT-Drude bath only as a function of the bath inverse temperature $\beta\hbar\omega_0$. The dashed and the solid black curves denote the RWA and the non-RWA couplings, respectively, $\gamma_B/\omega_0=0.2$ and $\Delta_B/\omega_0^{1/2}=0.4$. The dashed gray curve denotes steady states for decoherence in the stochastic field only, $\gamma_\nu/\omega_0=0.2$ and $\Delta_\nu/\omega_0^{3/2}=0.4$.	
}
\end{figure*}

With imaginary time HEOMs \cite{doi:10.1063/1.4890441} one obtains equilibrium states of the system by integrating over the imaginary time to the specified temperature. The HEOM (\ref{atom-bath-field-heom}) is a real time one, so the initial state has to be propagated (\ref{factorized-initial-conditions}) forward in time until the reduced density matrix cease changing. In Fig.~\ref{fig:steady-states} we illustrate steady states of an initially excited TLS interacting with different environments: an HT-Drude bath environment, a stochastic field environment, and the joint environment.

The stochastic field environment brings the TLS to a steady state where both the excited and the ground states are equally possible. In case of HT-Drude bath environment (the stochastic field is off), the steady states show no dependence on the bath frequency cutoff and the bath coupling strength. It resembles the case of interaction with the stochastic field environment, but with additional dependence on the bath temperature (see Fig.~\ref{fig:steady-states}(c)). In our model the temperature dependence is included perturbatively, so we can cover only a small range of temperatures near $\beta=0$. The excited state population tends to decrease with the inverse temperature $\beta$ and the slope of the line is greater for the RWA interaction Hamiltonian (\ref{HIB-RWA}). When $\beta$ decreases, the distance between the TLS steady states in the bath environment and the TLS steady states in the stochastic field environment becomes smaller. The Markovian approximation (\ref{markovian-me}) gives similar results.   

When the TLS interacts with the joint environment, we observe the interplay between decoherence channels corresponding to the subenvironments (Figs.~\ref{fig:steady-states}(a), \ref{fig:steady-states}(b)). Let us take frequency cutoffs of all the random processes constituting the stochastic field equal and change them simultaneously ($\gamma_{\nu}=\gamma_F$ for all $\nu$). In the beginning, when the stochastic field frequency cutoff $\gamma_F$ is zero, an increase of $\gamma_F$ causes an increase of the field contribution and an increase of the excited state probability of the steady state, because the steady state in the infinite temperature HT-Drude bath lies higher. Then we reach the maximum, and after it an increase of $\gamma_F$ leads to a decrease of the stochastic field contribution and a decrease of the TLS exited state population in the steady state (Fig.~\ref{fig:steady-states}(a)). In case of the RWA coupling with the bath the overall impact of the stochastic field is greater and makes the stationary states lie above the ones of the non-RWA case. By comparison, when the stochastic field is off, the RWA steady states are located below (Fig.~\ref{fig:steady-states}(c)). The Markovian approximation (\ref{markovian-me}) tends to wrongly overestimate the field contribution for small frequency cutoffs and gives more rapid decrease for bigger cutoffs, with no maximum in between, because the line starts from the steady state in the stochastic field environment. If we change the roles of the stochastic field and the bath and start manipulating the frequency cutoff of the bath $\gamma_B$, we obtain the inverted picture: no contribution of the bath at $\gamma_B=0$, followed by a minimum, and a region of constant growth after it. 

The observed dependence on the environment frequency cutoff can be explained by the form of the environment spectral density. Both the OU random process and the HT-Drude bath have spectrums with one peak. Because the spectrums have cutoffs in the algebraic form, the location of the peak depends on the frequency cutoff, and the peak moves to the right when the frequency cutoff increases, from $\omega=0$ to the TLS frequency $\omega=\omega_0$ and then further away. As a result, the impact of the environment has the maximum when the peak is located at the resonance $\omega=\omega_0$ and decreases in both directions from it. 
     
Now we take frequency cutoffs of all the random processes of the stochastic field constant and vary their coupling strengths in a similar way: we take them all equal $\Delta_{\nu}=\Delta_F$ and analyze the dependence on $\Delta_F$. The dependence of the TLS steady states on $\Delta_F$ is presented in Fig.~\ref{fig:steady-states}(b). In contrast to the case of dependence on the frequency cutoff $\gamma_F$, there are no maximums, because the spectrums of the random processes keep the same form for all coupling strengths, only their magnitude changes. At $\Delta_F=0$ the stochastic field is completely decoupled and we have the steady states of the TLS in the HT-Drude bath. When the coupling between the TLS and the stochastic field is strong, the stochastic field dominates the bath and the steady state approaches the steady state of the TLS in the stochastic field environment.         
     
A more detailed investigation of interaction with the stochastic field subenvironment, involving analysis of impact of changing the frequency cutoff and the coupling strength of a random process individually for each of the random processes of the stochastic field, reveals the differences originating from the different types of coupling of the random processes with the TLS. The random processes \(\xi_2(t)\) and \(\Omega(t)\) acting together with the HT-Drude bath subenvironment do not affect the TLS steady states in the Markovian approximation and the steady TLS excited state probability is always higher for the non-RWA interaction with the bath in comparison with the RWA interaction with the bath. In case of changing the random process \(\xi_1(t)\), the excited state probability behaves the same way only for small frequency cutoffs $\gamma_{\xi_1}$ and coupling strengths $\Delta_{\xi_1}$, otherwise the relation is reversed. If the stochastic field is represented by the \(\Omega(t)\) random process, it forms a purely dephasing channel but still impacts the steady states via the relaxation channel of the bath, as a consequence of the induced correlations between the subenvironments developed during the evolution.  

\section{\label{sec:dm-evol}Density matrix evolution}
\begin{figure}[t]
\includegraphics[width=\linewidth]{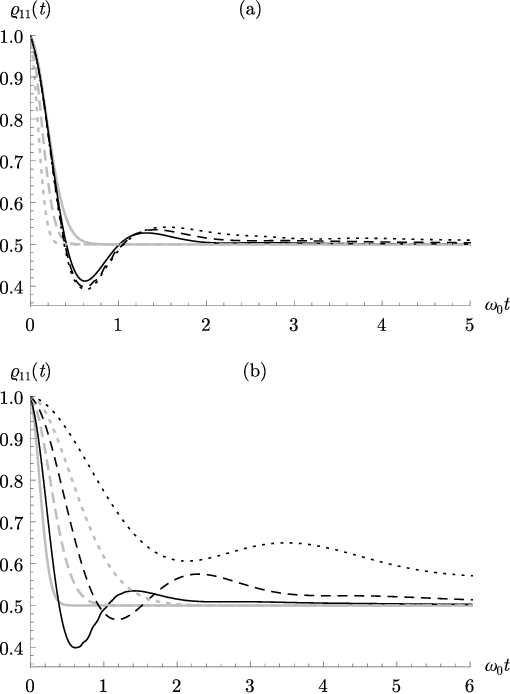}
\caption{
\label{fig:dm-evol-field}
Evolution of the TLS excited state population in the stochastic field in dependence on (a) the field frequency cutoff $\gamma_F$ and (b) the field coupling strength $\Delta_F$. Black denotes the non-
Markovian curves, gray stands for the Markovian ones. In (a) $\gamma_\nu/\omega_0=\gamma_F/\omega_0=(0.2, 0.4, 0.8)$ and $\Delta_\nu/\omega_0^{3/2}=\Delta_F^2/\omega_0^{3/2}=1.6$, for \{dotted, dashed, solid\} curves, respectively, and in (b) $\gamma_\nu/\omega_0=\gamma_F/\omega_0=0.4$, $\Delta_\nu/\omega_0^{3/2}=\Delta_F/\omega_0^{3/2}=\{0.4, 0.8, 1.6\}$. 
} 
\end{figure}

\begin{figure}[t]
\includegraphics[width=\linewidth]{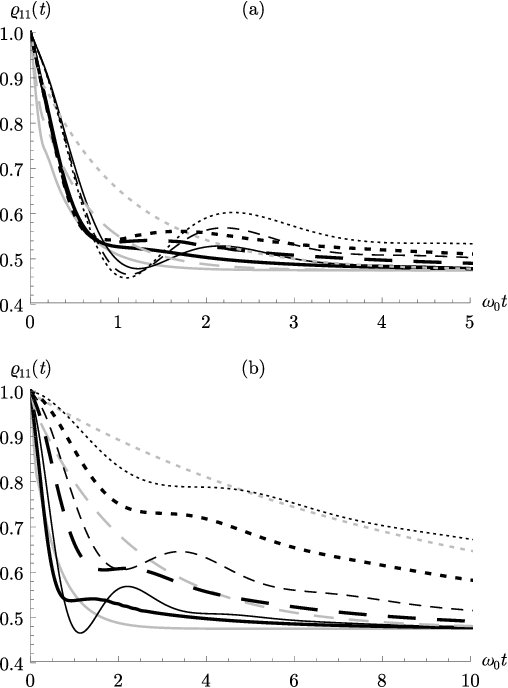}
\caption{
\label{fig:dm-evol-htdrude}
Evolution of the TLS excited state population in the HT-Drude bath (the stochastic field is off) in dependence on (a) the bath frequency cutoff $\gamma_B$ and (b) the bath coupling strength $\Delta_B$. The thin and the thick black curves (any stroke style) denote the RWA and the non-RWA couplings, respectively, the gray curves stand for the Markovian approximation. In (a) $\gamma_B/\omega_0=\{0.2, 0.4, 0.8\}$, $\Delta_B/\omega_0^{1/2}=1.6$, $\beta\hbar\omega_0=0.1$, for \{dotted, dashed, solid\} curves, respectively, and in (b) $\gamma_B/\omega_0=0.4\omega_0$, $\Delta_B/\omega_0^{1/2}=\{0.4, 0.8, 1.6\}$, $\beta\hbar\omega_0=0.1$.
}
\end{figure}

\begin{figure}[t]
\includegraphics[width=\linewidth]{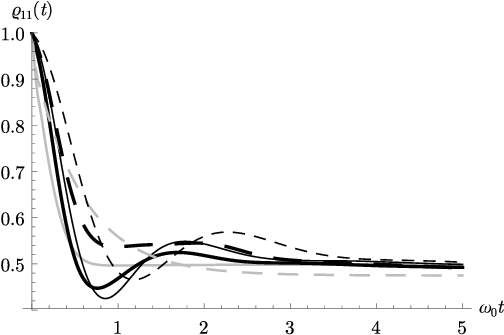}
\caption{
\label{fig:dm-evol-htdrude-field}
Evolution of the TLS excited state population for simultaneous interaction with the stochastic field and the HT-Drude bath (solid curves) in comparison with the case of interaction with the bath only (dashed curves). The thin and the thick black curves (any stroke style) denote the RWA and the non-RWA couplings, respectively, the gray curves stand for the Markovian approximation, $\gamma_\nu/\omega_0=\gamma_F/\omega_0=0.4$, $\Delta_\nu/\omega_0^{3/2}=\Delta_F/\omega_0^{3/2}=0.8$, and $\gamma_B/\omega_0=0.4$, $\Delta_B/\omega_0^{1/2}=1.6$, $\beta\hbar\omega_0=0.1$.
}
\end{figure}

Let us consider evolution of the initially excited TLS in the environments discussed in the previous section. In Fig.~\ref{fig:dm-evol-field} we show evolution of the excited state population in the stochastic field environment and in Fig.~\ref{fig:dm-evol-htdrude} we show the evolution in the HT-Drude bath environment for different frequency cutoffs and coupling strengths. In Fig.~\ref{fig:dm-evol-htdrude-field} the impact of the stochastic field on evolution of the TLS in the HT-Drude bath is shown. 

One specific feature of the evolution is the presence of rapidly vanishing oscillations. The oscillations are more noticeable for interaction with the bath environment in the RWA approximation (\ref{HIB-RWA}) and for interaction with the stochastic field environment. These oscillations can also be found for the non-RWA coupling with the bath (\ref{HIB-non-rwa}), but they are absent if the Markovian approximation (\ref{markovian-me}) is applied. The oscillations are a distinctive feature of a non-Markovian evolution and have a clear relation to the Rabi oscillations. The amplitude of the oscillations depends on the distance between the peak of the environment spectral density and the TLS resonance frequency. If the distance is large, there are no significant oscillations. It is the case of large frequency cutoffs and also of the Markovian approximation, because it implies interaction with a continuum of modes, which corresponds to large frequency cutoffs.

The character of the TLS excited state population evolution can be explained in a way similar to the one that we used in Sec.~\ref{sec:steady-states} for explanation of the TLS steady states behavior. Here, the impact of the environment depends on the location of the peak of the environment spectral function, determined by the environment frequency cutoff. An increase of the coupling strength speeds up the evolution, causing the steady state to be reached faster  (Figs.~\ref{fig:dm-evol-field}(b) and \ref{fig:dm-evol-htdrude}(b)). For the frequency cutoff the situation is more complex: for small cutoffs it takes more time to reach equilibrium same as for large cutoffs, while for moderate cutoffs it takes less time (Figs.~\ref{fig:dm-evol-field}(a) and \ref{fig:dm-evol-htdrude}(a)). Apparently it has close connections with the magnitude of the distance between the peak of the environment spectral density and the TLS resonance frequency.

For the TLS interacting with the HT-Drude bath environment, only one minimum and one maximum of the excited state population could be seen well for a wide range of frequency cutoffs \(\gamma_B\) (Fig.~\ref{fig:dm-evol-htdrude}(a)) and coupling strengths \(\Delta_B\) (Fig.~\ref{fig:dm-evol-htdrude}(b)). For small \(\gamma_B\) the difference between the minimum and the maximum is the biggest, and with growth of \(\gamma_B\) it gradually vanishes: the minimum slowly rises, the maximum lowers, but much faster, until they disappear; after this, the curve exhibits no oscillations. The main difference between evolution for the RWA coupling and the non-RWA coupling with the bath resides in the value of the first minimum. For strong RWA couplings it is deeper and drops lower than the stationary value, while for the non-RWA couplings it gradually vanishes when the coupling strength increases, remaining strictly above the stationary value. Because the maximum lowers too, the difference between the maximum and the minimum has a clear extremum. Overall, the non-RWA evolution resembles a smoothed version of the RWA evolution. The bath temperature influences the dynamics mainly by adjusting via the stationary states (Fig.~\ref{fig:steady-states}(c)).

Dynamics of the TLS in the stochastic field environment (Fig.~\ref{fig:dm-evol-field}) is similar to the dynamics in the infinite-temperature HT-Drude bath with the RWA coupling. We have the same dependence on frequency cutoffs and coupling strengths of random processes \(\xi_1(t)\) and \(\xi_2(t)\), but dependence on $\gamma_\Omega$ and $\Delta_\Omega$ is different. The Hamiltonian for interaction with $\Omega(t)$ commutes with the reduced density matrix and impacts the dynamics only in presence of another decoherence channel, e.g. one of the random processes \(\xi_1(t)\) and \(\xi_2(t)\) or the bath. The frequency cutoff $\gamma_\Omega$ mainly affects the steady state, an increase in the coupling strength slightly raises the minimum and practically does not affect the maximum.

In case of interaction with both the bath and the stochastic field (Fig.~\ref{fig:dm-evol-htdrude-field}), the presence of two subenvironments speeds up the evolution, i.e. equilibrium is reached earlier in comparison with interaction with only one of the subenvironments. The appearance of additional decoherence channels formed by the stochastic field increases the overall decoherence rate. The evolution resembles the one for interaction with the bath environment, but with a larger coupling strength. It can be explained by the similarities between the two environments, i.e. the stochastic field environment is similar to the infinite-temperature HT-Drude bath environment in terms of the impact on the TLS evolution. Another consequence of interaction with an additional stochastic field is the rise of the TLS excited state population in the steady state. The individual components of the stochastic field influence the evolution differently: an increase of \(\Delta_{\Omega}\) rises the minimum, an increase of \(\Delta_{\xi_2}\) lowers the minimum, and an increase of \(\Delta_{\xi_1}\) lowers the maximum. 

\section{\label{sec:spectrum}Emission spectrum}
Equilibrium emission spectrums of the TLS can be obtained by applying the Fourier transform to the two-time correlation function \(\langle\hat{\sigma}_+(t_2)\hat{\sigma}_-(t_1)\rangle\), where $t_2 > t_1$. The time $t_1$ is selected sufficiently big for the reduced density matrix evolution to reach its steady value. First the initial state is propagated to the steady state, then the operator \(\hat{\sigma}_-\) is applied to all the density matrices \(\hat{\rho}^{(A)}_{\bm{m}}(t_1)\), next the result is propagated to $t_2$, where \(\hat{\sigma}_+\) is applied. 

\begin{figure}[t]
\includegraphics[width=\linewidth]{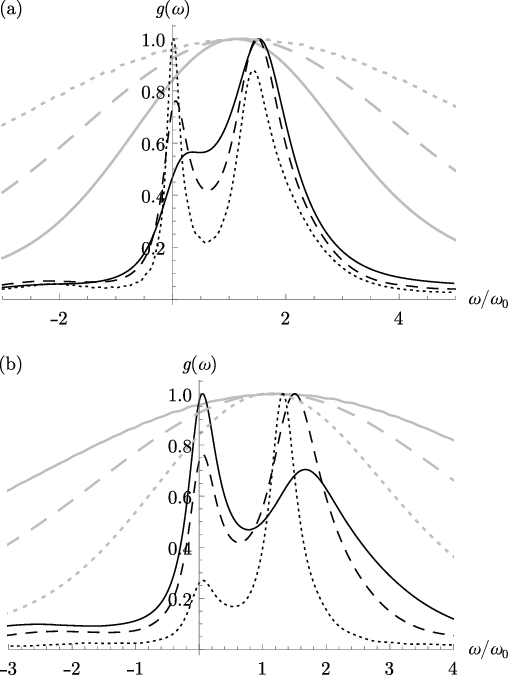}
\caption{
\label{fig:spectrum-field}
Emission spectrums of the TLS in the stochastic field, normalized by maximum values, in dependence on (a) the field frequency cutoff $\gamma_F$ and (b) the field coupling strength $\Delta_F$. Black denotes the non-Markovian curves, gray stands for the Markovian ones. In (a) $\gamma_F/\omega_0=\{0.1, 0.2, 0.4\}$, $\Delta_F/\omega_0^{3/2}=0.6$, for \{dotted, dashed, solid\} curves, respectively, and in (b) $\gamma_F/\omega_0=0.2$, $\Delta_F/\omega_0^{3/2}=\{0.4, 0.6, 0.8\}$.
}
\end{figure}
\begin{figure}[t]
\includegraphics[width=\linewidth]{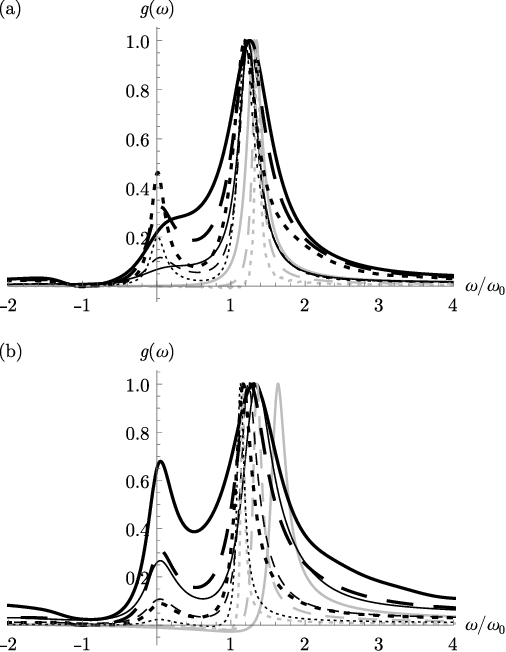}
\caption{
\label{fig:spectrum-htdrude}
Emission spectrums of the TLS in the HT-Drude bath, normalized by maximum values, in dependence on (a) the bath frequency cutoff $\gamma_B$ and (b) the bath coupling strength $\Delta_B$.  The thin and the thick black curves (any stroke style) denote the RWA and the non-RWA couplings, respectively, the gray curves stand for the Markovian approximation. In (a) $\gamma_B/\omega_0=\{0.1, 0.2, 0.4\}$, $\Delta_B/\omega_0^{1/2}=0.6$, $\beta\hbar\omega_0=0.1$, for \{dotted, dashed, solid\} curves, respectively, and in (b) $\gamma_B/\omega_0=0.2$, $\Delta_B/\omega_0^{1/2}=\{0.4, 0.6, 0.8\}$, $\beta\hbar\omega_0=0.1$.
}
\end{figure}

The emission spectrums for interaction with the HT-Drude bath environment (Fig.~\ref{fig:spectrum-htdrude}) are similar to the ones for interaction with the stochastic field environment (Fig.~\ref{fig:spectrum-field}). For large frequency cutoffs the peak is almost centered at \(\omega_0\), when the frequency cutoff falls, the peak shifts to the right and becomes less symmetrical, its left side rises faster than the right side, the peak widens while its intensity falls. At some frequency cutoff the peak at $\omega=0$ appears, becomes more distinct, rises, and, eventually, becomes the dominant peak, while the first one disappears. For some values of the coupling strength and the frequency cutoff, the side peak at $\omega=0$ is shifted to the right. It is more evident when the side peak appears for the first time and its intensity is close to the intensity of the main peak. This behavior is more typical for the non-RWA coupling, but it is also can be found for the RWA coupling with the bath. When the side peak appears with the intensity equal to the main peak intensity, the shift is the biggest. An increase of the coupling strength widens the main peak and slightly shifts it to the right. Because of the widening, the side peak cannot be seen clearly, but for large couplings it separates from the main one and then starts to dominate the spectrum, moving to the left and gradually approaching $\omega=0$. An increase of temperature makes the spectrum more asymmetrical, its left slope becomes less steep and the right slope becomes more steep. In the Markovian approximation the side peak never appears and the left slope of the main peak is much steeper than the right. Every non-RWA emission spectrum has a zero point located at \(\omega =-\omega_0\) for the set of parameters values studied. For some values of parameters both the Markovian approximation and the RWA lead to an erroneous absorption area between the main peak and the axes origin (Fig. \ref{fig:spectrum-htdrude}), which is absent if the full interaction Hamiltonian is used. 

\begin{figure}[tp]
\includegraphics[width=\linewidth]{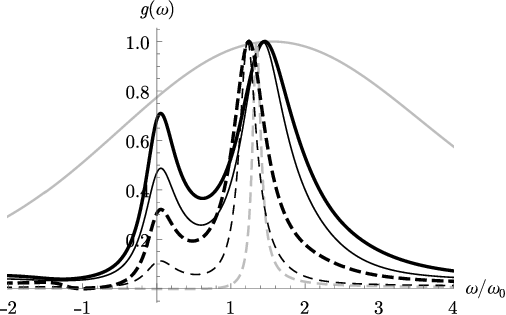}
\caption{
\label{fig:spectrum-htdrude-field}
Emission spectrums of the TLS for simultaneous interaction with the stochastic field and the HT-Drude bath (solid curves) in comparison with the case of interaction with the bath only (dashed curves), normalized by maximum values. The thin and the thick black curves (any stroke style) denote the RWA and the non-RWA couplings, respectively, the gray curves stand for the Markovian approximation, \(\gamma_{\nu }/\omega_0=\gamma_F/\omega_0=0.2\), \(\Delta_{\nu}/\omega_0^{3/2}=\Delta_F/\omega_0^{3/2}=0.4\), and $\gamma_B/\omega_0=0.2$, $\Delta_B/\omega_0^{1/2}=0.6$, $\beta\hbar\omega_0=0.1$.
}
\end{figure}

The impact of the stochastic field on the emission spectrums of the TLS in the bath environment is shown in Fig.~\ref{fig:spectrum-htdrude-field}. For the parameters values selected, the Markovian approximation is inaccurate both in absence and in presence of the stochastic field. In absence of the stochastic field it gives a much narrower contour and without the side peak on the left, in presence of the stochastic field the contour is excessively wide. The stochastic field smooths the negative frequency part of the spectrum and removes the zero point for the non-RWA coupling with the bath. The stochastic field effectively makes the interaction between the TLS and the bath stronger, which results in a higher side peak near $\omega=0$ and which widens the main peak and shifts it to the right.

\section{\label{sec:conclusions}Conclusions}
We have studied non-Markovian dynamics of a two level system interacting with two types of environments, a bosonic bath with the high-temperature Drude spectral density and a stochastic field of the Ornstein-Uhlenbeck type. By means of the influence functional approach we derived the hierarchical equations of motion \cite{doi:10.1143/JPSJ.75.082001} valid for arbitrary coupling strengths and environmental correlation times. The electric-dipole interaction between the two level system and the bath was considered in the exact form and in the rotating wave approximation. By means of the hierarchical equations of motion, we performed numerical analysis of the reduced density matrix evolution, steady states, and emission spectrums of the two level system for a wide range of the bath and the field frequency cutoffs and coupling strengths.  Also we studied the interplay between different decoherence channels of the joint environment combining the two subenvironments, the stochastic field and the bath, and analyzed the impact of the stochastic field on the evolution, steady states, and emission spectrums of the two level system in the bath environment. 
\bibliography{bibliography}% Produces the bibliography via BibTeX.

\end{document}